\documentclass[aps,pra,preprint,12pt]{revtex4-2}

\usepackage{amssymb}
\usepackage{amsmath}
\usepackage{bm}
\usepackage{natbib}
\usepackage{graphics,graphicx}
\usepackage{epsfig}
\usepackage{latexsym}
\usepackage{epic,eepic,graphicx,amssymb,amsmath,indentfirst}
\usepackage{bm}
\usepackage{graphicx}
\usepackage[T1]{fontenc}

\DeclareMathOperator{\sech}{sech}

\usepackage{multirow}

\usepackage{color}
\usepackage[colorlinks={true}]{hyperref}
\hypersetup{citecolor={blue}, filecolor={blue}, linkcolor={blue}, urlcolor={blue}}
\newcommand{\bra}[1]{\langle #1 \vert}
\newcommand{\ket}[1]{\vert #1 \rangle}

\begin{document}

\title{\bf Reverse engineering control of relative phase and populations of two-level quantum systems}

\author{Felipe Silveira Fagundes}
\author{Emanuel Fernandes de Lima}
\email{eflima@ufscar.br}
\affiliation{Departamento de F\'isica, Universidade Federal de S\~ao Carlos (UFSCar)\\ S\~ao Carlos, SP 13565-905, Brazil}
\date{\today}

\begin{abstract}
We consider the simultaneous control of the relative phase and populations of two-level quantum systems by an external field. We apply a reverse engineering approach, which allows obtaining an analytical expression for the control field depending upon two user-defined functions that dictate the population and the relative phase dynamics. We show that, in general, the prescribed functions for the dynamics cannot be chosen arbitrarily. We implement the reverse engineering technique to reach several target states using different kinds of functions to specify the system dynamics. We show that by adjusting these dynamical functions, we can produce different kinds of control fields. These controls can be easily build, needing, apart from the dynamical function themselves, only their derivatives. The methodology presented here will certainly find many applications that go beyond simple two-level systems. 
\end{abstract}

\maketitle

\newpage

\section{Introduction}

The investigation of two-level quantum systems is of fundamental interest for Physics \cite{Li2019,10.1063/5.0188749,PhysRevApplied.13.024066,PhysRevA.94.063826}. Despite their simplicity, two-level systems can represent to a very good approximation diverse practical situations and have proven to be adequate to understand the physics content of a variety of coherent phenomena. For instance, a two-level quantum system that interacts with an external time-dependent field can represent an atom under the action of a laser field \cite{eberly}. The study of two-level systems is also crucial for the development of new technologies: quantum computing faces this kind of system as the fundamental unit of information, the \textit{qubit} \cite{Bennett2000,doi:10.1063/1.1818131,MISHIMA201063}. Thus, it is not a surprise that the search for analytical solutions of driven two-level systems, which started since the early days of quantum mechanics, remains to the present days \cite{PhysRevLett.109.060401}. These solutions provide deep understanding of the dynamics and also plays essential role in applications.   

A particularly significant problem in most of the recent applications is the control of the two-level system, i.e., the search of control fields that can drive the initial state to a desired target state at some finite time \cite{e25020212}. The special case of population inversion, where the population of the levels are swept at the final time, can be achieved by several means, e.g., using $\pi$-pulses \cite{10.1063/1.481993}, chirped pulses \cite{PhysRevLett.80.1406}, or adiabatic passage \cite{Demirplak2003}, to name a few. A more general approach is provided by the optimal quantum control framework, which allows to drive any initial state to an arbitrary final state \cite{10.1063/1.458438,PhysRevLett.125.250403,doi:https://doi.org/10.1002/9783527639700.ch5}. The optimal quantum control equations are usually solved numerically and one is often interested in reaching the desired state regardless of the detailed system dynamics.

In an alternative procedure, often referred to as a reverse engineering technique, the control field is designed to follow a dynamical constraint \cite{Zhang2017,PhysRevApplied.8.054008,Wang2024,Vitanov_2015}. This methodology is encompassed by quantum tracking control, which seeks to finds control fields to drive an expectation value of a observable along a prescribed time-dependent \textit{track} \cite{Mirrahimi2005,PhysRevA.108.033106,PhysRevA.98.043429,CHEN19971617,PhysRevA.47.4593}. This is performed by inverting the equations of motion in order to solve for the control field. Such approach avoids iterative optimization and thus can be computationally less demanding but suffers from the potential appearance of singularities presented in the inversion of the dynamical equations.

The reverse-engineering control of the population dynamics of two-level systems by a resonant pulse has been proposed in Ref. \cite{articleGolubev}. In this work, an analytical expression for the control field was furnished as a function of an \textit{a priori} specified population dynamics. Apart from the rotating wave approximation, the phases of the dynamical coefficients were assumed to be fixed, which restricted the applicability of the deduced formula. This work has been followed by investigations considering two-level system under the influence of dissipative effects \cite{PhysRevA.101.023822,GRIRA2021104419,PhysRevA.100.012103}. More recently \cite{articleAndras}, the problem of controlling both the populations and phases of two-level quantum system has been tackled, extending the formula derived in Ref. \cite{articleGolubev}. In this approach, the author defines two functions, one for the population and other for the phase of one of the coefficients of the quantum levels. From an analytical integration, the author obtained the phase of the remaining coefficient and then deduced an analytical formula for the control field. One drawback of this approach is the need of performing an integration analytically, which, even when it is possible, in general yields a quite complicate analytical formula for the control field.

In the present work, we address the problem of the simultaneous control of the population and the relative phase of two-level quantum systems through the reversing engineering technique, thus extending previous results \cite{articleGolubev,articleAndras}. The expression of the control field obtained by inverting the equations of motion depends only on the population, the relative phase and their time derivatives. As a consequence, the control field is determined by prescribing two function for each one of these quantities. However, additional constraints also follow from the dynamical equations, which implies that the dynamical functions for the population and relative phase cannot be chosen completely arbitrarily. We show that by appropriately choosing these dynamical functions, we obtain analytical expression for the control field without the need of performing any integration. Several initial and target states are chosen to illustrate the approach.

\section{The control framework}

We consider a two-level system interacting with a linearly-polarized, time-dependent external field acting from an initial time $t=t_0$ to a final time $t=t_f$. This system can model, for instance, a two-level atom interacting with a laser pulse. From the knowledge of the initial conditions, our goal will be to specify, as much as possible, the system dynamics until a desired target final state is reached. We remark that the present development follows somehow closely Refs.\cite{articleGolubev,articleAndras}, but with some key differences. We denote the ground level by $\ket{g}$, with energy $E_g$, and the excited level by $\ket{e}$, with energy $E_e$. In atomic units, we write the corresponding total time-dependent Hamiltonian as,

\begin{equation}\label{TH}
    H=H_0-\mu \varepsilon(t)\left(\ket{g}\bra{e}+\ket{e}\bra{g}\right),
\end{equation}
with $H_0$ being the unperturbed Hamiltonian of the two-level system,
\begin{equation}\label{H0}
H_0=E_g\ket{g}\bra{g}+E_e\ket{e}\bra{e},
\end{equation}
and $\mu$ the projection of the electric dipole moment along the field polarization axis. The time-dependent function $\varepsilon(t)$ stands for the electric field and plays the role of the control function. This external field, assumed to be real and with a carrier frequency $\omega$, can be expressed for convenience as two complex-conjugated parts,

\begin{equation}\label{Efield}
    \varepsilon(t)=\epsilon(t){\rm e}^{-{\rm i}\omega t}+\epsilon^*(t){\rm e}^{{\rm i}\omega t},
\end{equation}
where the complex function $\epsilon(t)$ ultimately defines the envelope and amplitude of the field and the asterisk denotes the complex conjugate.

The wavefunction can be written in the basis of the eigenstates of the unperturbed Hamiltonian as,

\begin{equation}
    \ket{\psi(t)}=C_g(t){\rm e}^{-{\rm i}E_g t}\ket{g}+C_e(t){\rm e}^{-{\rm i}E_e t}\ket{e},
\end{equation}
where the time-dependent coefficients $C_g(t)$ and $C_e(t)$ are the complex amplitudes of the ground and excited levels, respectively, in the interaction picture. Upon substitution of this expansion and Eq.~(\ref{Efield}) in the time-dependent Schrödinger equation, while invoking the rotating wave approximation (RWA), we obtain the following coupled system of differential equations for the coefficients,

\begin{align}\label{coupSys}
    \dot{C_g}(t) & =  {\rm i}C_e(t) \mu \epsilon^*(t) {\rm e}^{{\rm i}\left(\omega-\omega_0\right) t}  \nonumber \\
    \dot{C_e}(t) & =  {\rm i}C_g(t) \mu \epsilon(t) {\rm e}^{-{\rm i}\left(\omega-\omega_0\right) t}
\end{align}
where $\omega_0=E_e-E_g$ is the resonance frequency between the energy levels and the dots denotes the time derivative.

We can express each complex coefficient in terms of a time-dependent amplitude and phase,
\begin{equation}\label{coeff}
    C_j(t)=c_j(t){\rm e}^{{\rm i} \phi_j(t)},
\end{equation}
with the index $j$ denoting either the level $g$ or $e$, $c_j(t)=| C_j(t) |$ the absolute value and $\phi_j(t)$ the phase of the corresponding complex coefficient.

Our goal in the work is to derive control fields that yields a prescribed system dynamics. To this end, we rewrite each one of the equations in (\ref{coupSys}) in terms of $\epsilon(t)$ and $\epsilon^*(t)$ using Eq.(\ref{coeff}),

\begin{align}\label{field_comp}
\epsilon^*(t) & = -\frac{{\rm i}}{\mu}\frac{\dot{c}_g(t)+{\rm i}\dot{\phi}_g(t) c_g(t)}{c_e(t)} {\rm e}^{-{\rm i}\left[\left(\omega-\omega_0\right) t-\phi(t)\right]} \nonumber \\
\epsilon(t) & = -\frac{{\rm i}}{\mu}  \frac{\dot{c}_e(t)+{\rm i}\dot{\phi}_e(t) c_e(t)}{c_g(t)} {\rm e}^{{\rm i}\left[\left(\omega-\omega_0\right) t-\phi(t)\right]}
\end{align}
where we have defined the relative phase between the coefficients as $\phi(t)=\phi_g(t)-\phi_e(t)$. Substituting Eqs.~(\ref{field_comp}) in Eq.~(\ref{Efield}) and using the conservation of the norm,  $c_g\dot{c}_g=-c_e\dot{c}_e$,

\begin{align}\label{fielddev1}
    \varepsilon(t)=-\frac{{\rm i}}{\mu}\left[\left(\frac{\dot{c}_g(t)}{c_e(t)}+\frac{{\rm i}\dot{\phi}_g(t) c_g(t)}{c_e(t)} \right) {\rm e}^{{\rm i}\left[\omega_0t+\phi(t)\right]} +
     \left(-\frac{\dot{c}_g(t)}{c_e(t)}+\frac{{\rm i}\dot{\phi}_e(t) c_e(t)}{c_g(t)}\right) {\rm e}^{-{\rm i}\left[\omega_0t+\phi(t)\right]}\right].
\end{align}
Since we have assumed that the field is a real function, we obtain the following relation among the time derivatives of the phases,
\begin{equation}\label{phase_rel}
    \dot{\phi}_e(t)=\frac{c_g(t)^2}{c_e(t)^2}\dot{\phi}_g(t)=\frac{c_g(t)^2}{c_e(t)^2-c_g(t)^2}\dot{\phi}(t).
\end{equation}
An important point revealed by the above equation is that when the populations of the levels are equal to each other, the derivative of the relative phase has to vanish.

Equation ~(\ref{phase_rel}) and the fact that $c_e(t)^2+c_g(t)^2=1$ allow us to write the control field in Eq.~(\ref{fielddev1}) as,
\begin{align}\label{fielddev2}
    \varepsilon(t)=\frac{2}{\mu}\frac{\dot{c}_g(t)}{|\dot{c}_g(t)|\sqrt{1-c_g(t)^2}} \sqrt{\dot{c}_g(t)^2+\dot{\phi}_g(t)^2 c_g(t)^2}\sin{\left[\omega_0t+\phi(t)+\lambda(t)\right]},
\end{align}
where the phase $\lambda(t)$ is given by,
\begin{equation}
    \lambda(t)=\arctan{\frac{\dot{\phi}_g(t) c_g(t)}{\dot{c}_g(t)}}=\arctan{\frac{1-c_g(t)^2}{1-2c_g(t)^2}\frac{c_g(t)}{\dot{c}_g(t)}}\dot{\phi}(t).
\end{equation}

We are now in a position to apply the reverse engineering idea to our control problem. The dynamical equations (\ref{coupSys}) are to be accompanied by a set of initial conditions specifying the initial populations, $c_g(t_0)^2\equiv P^i$, $c_e(t_0)^2=1-c_g(t_0)^2$, and the initial phases $\phi_g(t_0)$, $\phi_e(t_0)$ and $\phi(t_0)=\phi_g(t_0)-\phi_e(t_0)\equiv\Phi^i$. From these initial conditions, we set up the population dynamics of the levels and also of the relative phase until they reach the desired target values $c_g(t_f)^2\equiv P^f$ and $\phi(t_f)\equiv\Phi^f$. To accomplish this task, we define two dynamical functions $P(t)$ and $\Phi(t)$ as

\begin{align}
    c_g(t) & =\sqrt{P(t)} \nonumber \\
    \phi(t) & =\Phi(t),
\end{align}
such that they match the initial and final conditions:  $P(t_0)=P^i$, $\Phi(t_0)=\Phi^i$, $P(t_f)=P^f$ and  $\Phi(t_f)=\Phi^f$. 

From Eq.~(\ref{fielddev2}), the engineered external field can be expressed in terms of the dynamical functions $P(t)$ and $\Phi(t)$,

\begin{align}\label{Controlfield}
    \varepsilon(t)=V_0(t) \sin{\left[\omega_0t+\Phi(t)+\Lambda(t)\right]},
\end{align}
where the amplitude and the phase are written as

\begin{align}\label{Bt}
    V_0(t) & =\frac{2}{\mu}\frac{\dot{P}(t)}{|\dot{P}(t)|}\left[\frac{\dot{P}(t)^2}{4P(t)(1-P(t))}+\frac{\dot{\Phi}(t)^2 P(t)(1-P(t))}{(1-2P(t))^2}\right]^{1/2} \\
    \Lambda(t) & =\arctan\left[\frac{P(t)(1-P(t))}{\dot{P}(t)}\frac{2\dot{\Phi}(t)}{1-2P(t)}\right].\label{lambda}
\end{align}
Note that there will be no singularity in expressions (\ref{Bt}) and (\ref{lambda}) \textit{if} $\dot{\Phi}(t)=0$ when $P(t)=1/2$ and $\dot{P}(t)=0$ in the extrema $P=1$ and $P=0$.

Therefore, in the present approach, two dynamical functions $P(t)$ and $\Phi(t)$ are chosen so that they match the initial conditions and reach the desired values of population and relative phase at the final time. The control field that yields the desired dynamics is then given by Eq.~(\ref{Controlfield}). However, in addition to the expected smoothness of the dynamical functions, they have to satisfy others constraints: (i) $P(t)$ has to take into account the conservation of the norm $c_g(t)^2+c_e(t)^2=1$, so that $0\leq P(t)\leq1$; (ii) $\dot{P}(t)=0$ whenever the population reach an extremum, i.e., $P(t)=0$ or $P(t)=1$; (iii) As noticed in Eq.~(\ref{phase_rel}), the derivative of the relative phase has to vanish whenever the populations are equal, that is,  $\dot{\Phi}(\tilde{t})=0$ for all $\tilde{t}$ such that $P(\tilde{t})=1/2$; (iv) We must have in mind that we are under the RWA, so that abrupt changes in the functions compared to the system natural period can break the approximation. Despite these constraints, there is still enough room for choosing the functions to govern the dynamics, as we shall see in the next section.

\begin{figure}[t]
\includegraphics[width=15cm]{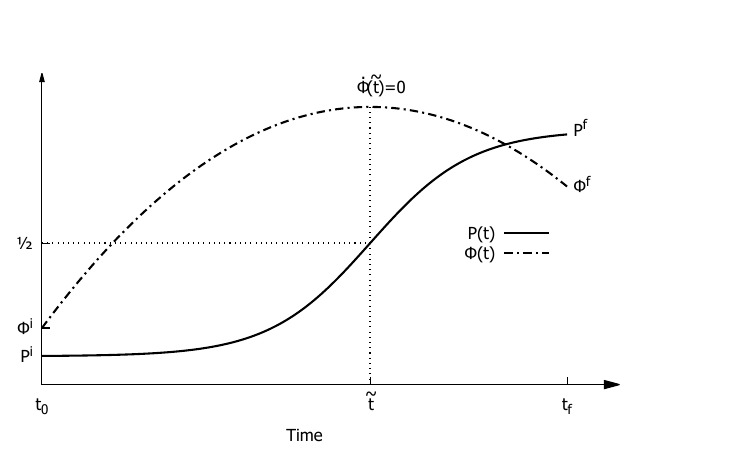}
\centering
 \caption{Schematic drawing of possible dynamical functions $P(t)$ and $\Phi(t)$. The system is driven from $P(t_0)=P^i$ and $\Phi(t_0)=\Phi^i$ to the target values $P(t_f)=P^f$ and $\Phi(t_f)=\Phi^f$. Note that $\Phi(t)$ is chosen such that $\dot{\Phi}(\tilde{t})=0$, while $\tilde{t}$ is defined by the condition $P(\tilde{t})=1/2$.}
\label{fig:scheme}
\end{figure}

Figure~\ref{fig:scheme} depict the possible choice of the dynamical functions $P(t)$ and $\Phi(t)$. In this case, it is intended to drive the system from $P(t_0)=P^i$ and $\Phi(t_0)=\Phi^i$ to the target values of population $P(t_f)=P^f$ and relative phase  $\Phi(t_f)=\Phi^f$. The dynamical functions are chosen in order to satisfy the constraints. In special, it is highlighted the fact that the time derivative of the relative phase is zero when the level populations are equal. Note that a myriad of possible functions could be chosen and that the value of $\tilde{t}$ is defined by the selection of $P(t)$.

Finally, it is worth considering two limiting cases. First, if the relative phase is intended to be constant, $\dot{\Phi}(t)=0$ in the hole interval, and we obtain for the control field,
\begin{align}\label{CFphip0}
    \varepsilon_{\Phi}(t)= \frac{1}{\mu}\frac{\dot{P}(t)}{\sqrt{P(t)(1-P(t))}} \sin{\left[\omega_0t+\Phi\right]},
\end{align}
and this is the result obtained in Ref.~\cite{articleGolubev}. Now, if the population is fixed, $\dot{P}(t)=0$ in the hole interval and the control field can be expressed as,

\begin{align}\label{CFPp0}
    \varepsilon_{P}(t)= \frac{2}{\mu}\frac{\dot{\Phi}(t)\sqrt{P(1-P)}}{1-2P} \cos{\left[\omega_0t+\Phi(t)\right]}.
\end{align}
Note that Eq.~(\ref{CFPp0}) will not work if the populations are to be kept in $1/2$, because is not possible to change the relative phase while keeping equal populations, see Eq.~(\ref{phase_rel}). In fact, Eq.~(\ref{CFPp0}) should be avoided if the population is to be kept too close to $1/2$ due to the small denominator.

\section{Simultaneous control of relative phase and populations}

Here, we illustrate several cases for the choice of the dynamical functions $ P(t) $ and  $ \Phi(t)$ for different objectives. Following Ref.~\cite{articleAndras}, the parameters of the two level system are chosen to model the $3{\rm s}\rightarrow 3{\rm p}$ atomic transition of sodium, for which $\omega_0=2.1$eV and $\mu=2.479$ atomic units. In all cases, we set $t_0=0$. The engineered control field is used to solve numerically the Schrödinger equation without the RWA.
 
\begin{figure}[h]
\includegraphics[width=8cm]{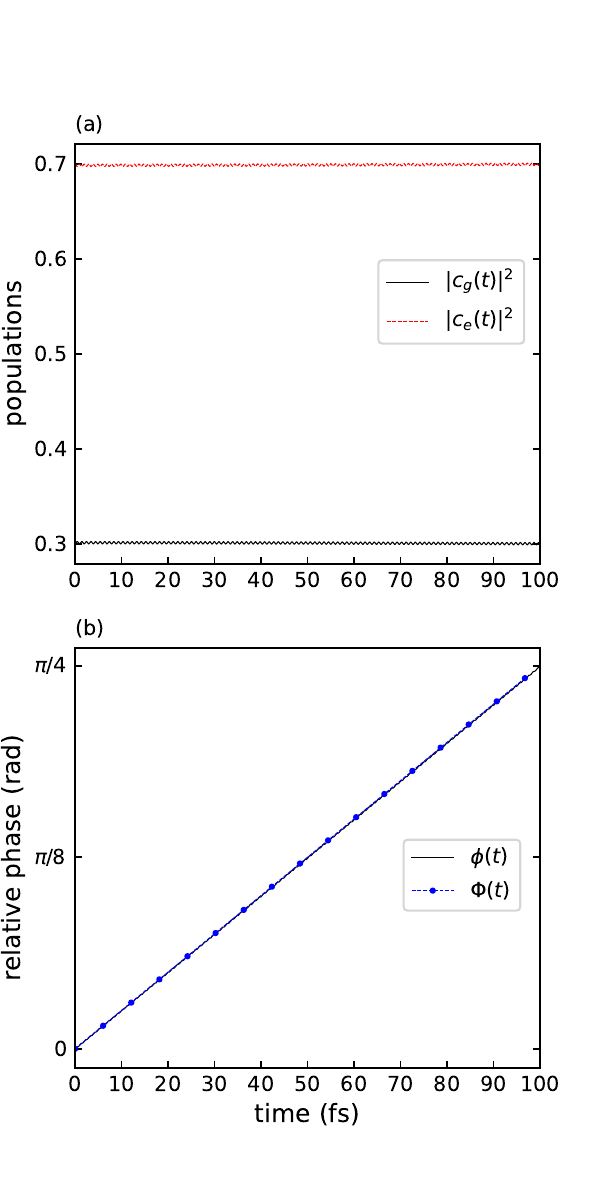}
\centering
 \caption{$P(t)$ is a constant at $P^i=P^f=0.3$. The initial relative phase is $\Phi^i=0$ and the target phase is $\Phi^f=\pi/4$. A linear polynomial is chosen for the dynamical function $\Phi(t)$ and formula (\ref{CFPp0}) is used. (a) Levels population dynamics $|c_j(t)|^2$ and (b) Relative phase dynamics $\phi(t)$ (continuous line) compared to the chosen dynamics function $\Phi(t)$ (dotted line with points).}
\label{fig:LC2}
\end{figure}

Initially, we consider the situation in which one desires to change only the relative phase, while keeping the populations constants during the hole time interval. We assume the population of the ground level is to be fixed in $0.3$ and it is desired a change of the relative phase from $\Phi^i=0$ to $\Phi^f=\pi/4$, with $t_f=100$fs. Thus, the function $P(t)$ is just a constant equal to $P^i=P^f=0.3$ and we chose $\Phi(t)$ as a straight line joining $\Phi^i$ and $\Phi^f$. We then apply formula (\ref{CFPp0}) for the control field. Note that in this case, the control field is a square pulse with constant amplitude of $0.9\times10^{8}{\rm V/m}$ and blue-shifted from the resonance frequency $\omega_0$ by $5.17$meV. Figure~\ref{fig:LC2} shows the numerical results obtained. Panel (a) shows that the populations are essentially constants, despite small fast oscillations. The desired relative phase is also reached as shown in panel (b). Similar good results are obtained for different values of the target relative phase, maintaining the other parameters fixed. But for larger values of $\Phi^f$ the oscillations in the populations increases. This is expected since the amplitude of the control field depends on $\dot{\Phi}(t)$, so the larger the changes in the relative phase the larger the control field amplitude. However, formula~(\ref{CFPp0}) is not practical for the populations close to $1/2$ due to the singularity at $P(t)=1/2$. In this case, Eq.~(\ref{Controlfield}) should to be used instead and, though we can set $P^i=P^f=1/2$, the population cannot be fixed at $1/2$ during the hole interval. We will consider this situation latter on in this Section.

\begin{figure}[h]
\includegraphics[width=15cm]{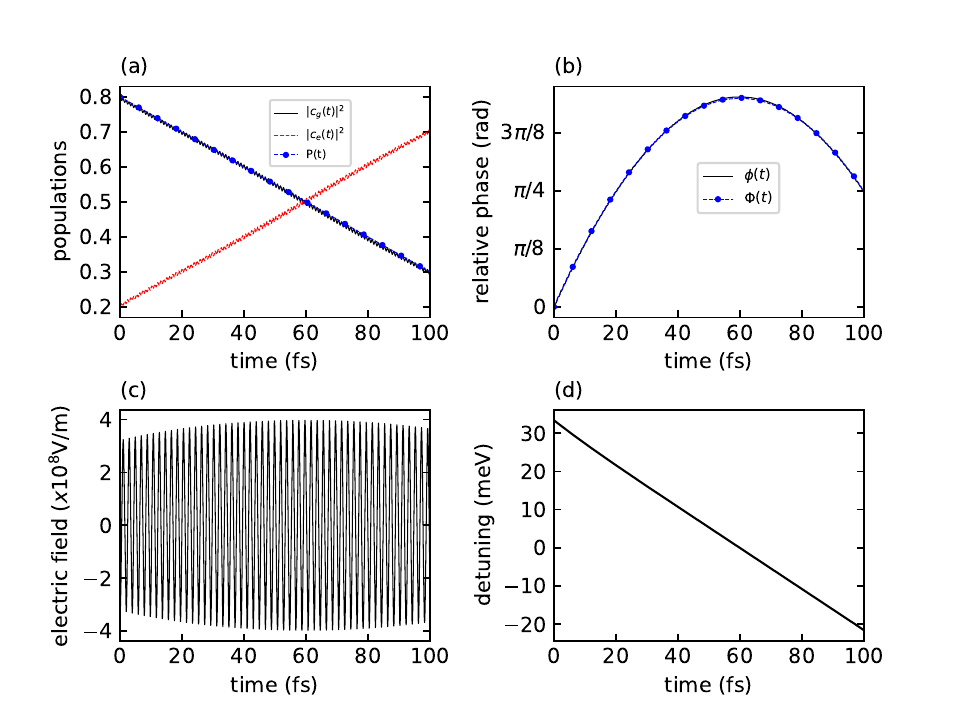}
\centering
 \caption{$P(t)$ is a linear function joining $P^i=0.8$ to $P^f=0.3$. The initial relative phase is $\Phi^i=0$ and the target phase is $\Phi^f=\pi/4$. A quadratic polynomial is chosen for the relative phase such that $\Phi(t_0)=\Phi^i$, $\Phi(t_f)=\Phi^f$ and $\dot{\Phi}(\tilde{t})=0$, where $\tilde{t}=60$fs. (a) Levels population dynamics along with $P(t)$; (b) Relative phase dynamics $\phi(t)$ (continuous line) compared to the chosen dynamics function $\Phi(t)$ (dotted line with points); (c) Applied electric field, Eq.~(\ref{Controlfield}), and (d) Detuning from the resonance frequency $\dot{\Phi}(t)+\dot{\Lambda}(t)$.}
\label{fig:LinpopQuadphase}
\end{figure}

Now consider a linear function for the populations $P(t)$ and a quadratic function for the relative phase $\Phi(t)$. We again set $t_f=100$fs. Assume it is desired to drive the population of the ground level from $P^i=0.8$ to $P^f=0.3$, while the relative phase should go from $\Phi^i=0$ to $\Phi^f=\pi/4$. Figure \ref{fig:LinpopQuadphase} shows the results of applying formula (\ref{Controlfield}) to this case. Note that since $P(t)$ is a linear function, the time at which $P(t)=1/2$ is $\tilde{t}=60$fs. Thus, $\Phi(t)$ is build to match the initial and final phases and also $\dot{\Phi}(\tilde{t})=0$. The corresponding results are shown in Fig.~(\ref{fig:LinpopQuadphase}). In panel (a), apart from small oscillations of the populations, the dynamics follows reasonably the  prescribed linear behavior. The dynamics of the relative phase is quadratic with a maximum at $t=\tilde{t}$ as shown in panel (b). Panel (c) shows that the control field has an almost constant envelope. In panel (d) it is shown the detuning $\dot{\Phi}(t)+\dot{\Lambda}(t)$, which represents the deviation of the field frequency from $\omega_0$. This panel shows that the control field is essentially a linear chirped pulse. This scheme works reasonably well for several target values of population and phase. It should be noted however that the choice of a linear function for the population dynamics preclude $P^i$ or $P^f$ of being equal to zero or one, because, as already mentioned, $\dot{P}(t)$ should be zero at the extrema. Also note that $\Phi(t)$ cannot be chosen as a quadratic function if $\Phi^i\neq \Phi^f$ and $\tilde{t}$ is exactly in the middle of the time interval, because $\Phi(t)$ could not interpolate the initial and target values while satisfying $\dot{\Phi}(\tilde{t})=0$.

\begin{figure}[h]
\includegraphics[width=15cm]{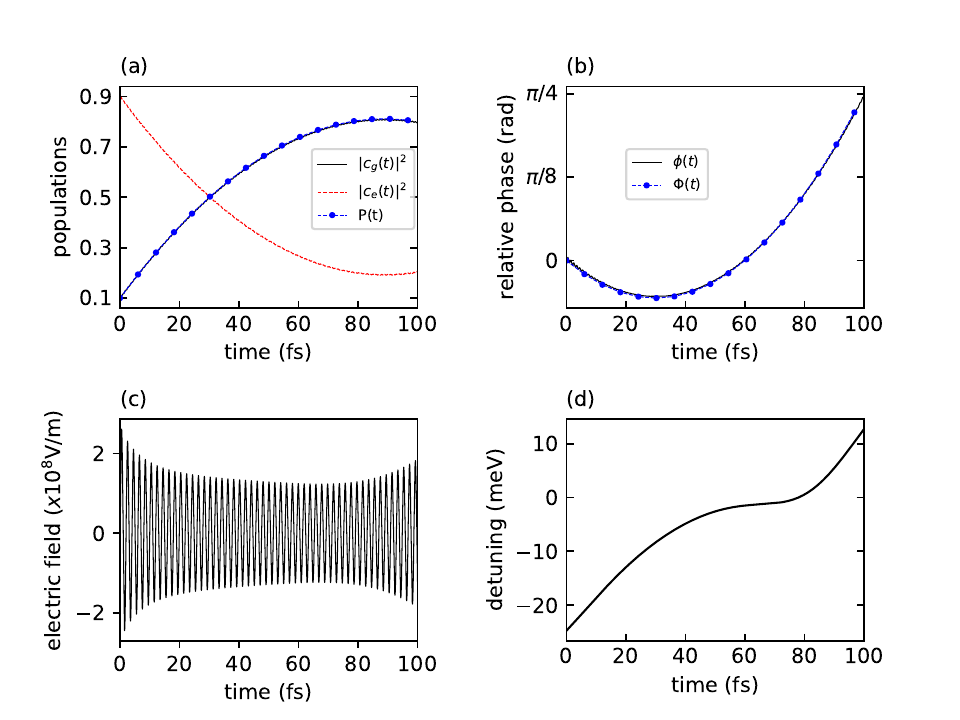}
\centering
 \caption{$P(t)$ is a quadratic polynomial joining $P^i=0.1$ to $P^f=0.8$ and such that $P(\tilde{t})=1/2$ for $\tilde{t}=30$fs. The initial relative phase is $\Phi^i=0$ and the target phase is $\Phi^f=\pi/4$. A quadratic polynomial is chosen for the relative phase such that $\Phi(t_0)=\Phi^i$, $\Phi(t_f)=\Phi^f$ and $\dot{\Phi}(\tilde{t})=0$. (a) Levels population dynamics along with $P(t)$; (b) Relative phase dynamics $\phi(t)$ (continuous line) compared to the chosen dynamics function $\Phi(t)$ (dotted line with points); (c) Applied electric field, Eq.~(\ref{Controlfield}), and (d) Detuning from the resonance frequency $\dot{\Phi}(t)+\dot{\Lambda}(t)$.}
\label{fig:QuadpopQuadphase}
\end{figure}

Let us examine the choice of $P(t)$ as a quadratic polynomial. Since $P(t)$ has to connect the initial population $P^i$ to the target population $P^f$, there is an additional point that can be arbitrarily chosen to specify $P(t)$. This extra point may be used to set the value of $\tilde{t}$ for which $P(\tilde{t})=1/2$ and $\dot{\Phi}(\tilde{t})=0$. Figure \ref{fig:QuadpopQuadphase} shows the results for this case, where $\Phi(t)$ is also a quadratic polynomial and $t_f=100$fs. We set $P^i=0.1$, $P^f=0.8$, $\Phi^i=0$ and $\Phi^f=\pi/4$. In panel (a), apart from imperceptible oscillations of the populations, the dynamics follows reasonably the  intended quadratic behavior. Panel (b) shows the dynamics of the relative phase with a minimum at $t=\tilde{t}$. Panel (c) shows that the control field decreases its amplitude roughly until $t=20$fs, then stays approximately constant up to $t=80$fs, and increases its amplitude from $t=80$fs. Panel (d) shows that the detuning is roughly constant in the interval $t=50$fs to $t=80$fs, and increases approximately linearly from $0$ to $t=50$fs and from $t=80$fs to $t=100$fs. This choice for the dynamical functions works very well for several target values of population and phase. However, despite the freedom to set the value of $\tilde{t}$, there is still a drawback if the values of $P^i$ or $P^f$ are either zero or one, as in the previous case, since we cannot generally set $\dot{P}(t)=0$ at the extrema.

\begin{figure}[h]
\includegraphics[width=15cm]{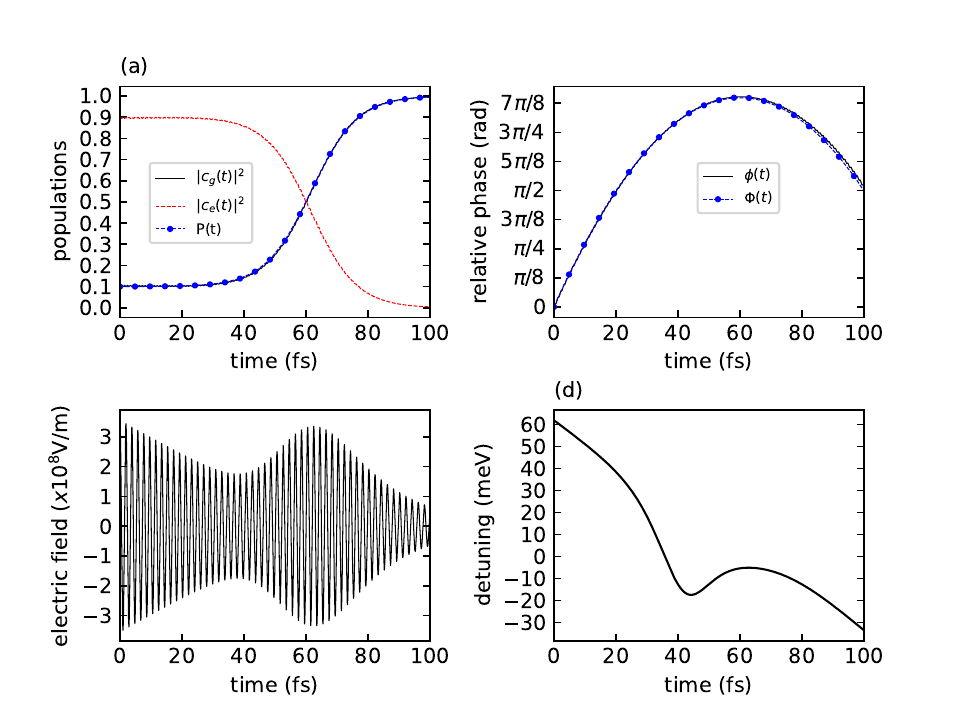}
\centering
 \caption{$P(t)$ is a hyperbolic tangent joining asymptotically $P^i=0.1$ to $P^f=1$ and such that $P(\tilde{t})=1/2$ for $\tilde{t}=60$fs, see Eq.~(\ref{Ptanh}). The initial relative phase is $\Phi^i=0$ and the target phase is $\Phi^f=\pi/2$. A quadratic polynomial is chosen for the relative phase such that $\Phi(t_0)=\Phi^i$, $\Phi(t_f)=\Phi^f$ and $\dot{\Phi}(\tilde{t})=0$. (a) Levels population dynamics along with $P(t)$; (b) Relative phase dynamics $\phi(t)$ (continuous line) compared to the chosen dynamics function $\Phi(t)$ (dotted line with points); (c) Applied electric field, Eq.~(\ref{Controlfield}), and (d) Detuning from the resonance frequency $\dot{\Phi}(t)+\dot{\Lambda}(t)$.}
\label{fig:TanhpopQuadphase}
\end{figure}

We consider now the choice of $P(t)$ as a hyperbolic tangent function given by,
\begin{equation}\label{Ptanh}
  P(t)=A\tanh(\alpha t+\beta)+B,
\end{equation}
where $A=(P^f-P^i)/2$, $B=(P^f+P^i)/2$, and 
$$
\beta=\frac{1}{2}\ln\frac{1+\gamma}{1-\gamma}-\alpha \tilde{t},
$$
where $\gamma=(1/2-B)/A$. With this setup, $P(t)$ approaches asymptotically the initial and the target values of the populations $P^i$ and $P^f$, while satisfying $P(\tilde{t})=1/2$. As in the case of the quadratic polynomial, the hyperbolic function allows to tune the value of $\tilde{t}$, but in addition the derivatives of $P(t)$ at the initial and final times can be approximately zero. The parameter $\alpha > 0$ sets how fast is the transition rate between the levels, which occurs around $\tilde{t}$. Figure \ref{fig:TanhpopQuadphase} shows a specific situation, where we have set $P^i=0.1$, $P^f=1$, $\Phi^i=0$ and $\Phi^f=\pi/2$. Once again, $t_f=100$fs. Additionally, we have selected $\alpha=0.068{\rm fs}^{-1}$ and $\tilde{t}=60$fs. From panels (a) and (b), we note that the population and the relative phase follows the prescribed dynamical functions. Panels (a) to (c) show that up to $t=40$fs the field acts to change the relative phase and only then starts to effectively change the populations. The detuning shown in panel (d) has a marked depression around $t=45$fs combined with a general approximately linear decreasing.

\begin{figure}[h]
\includegraphics[width=15cm]{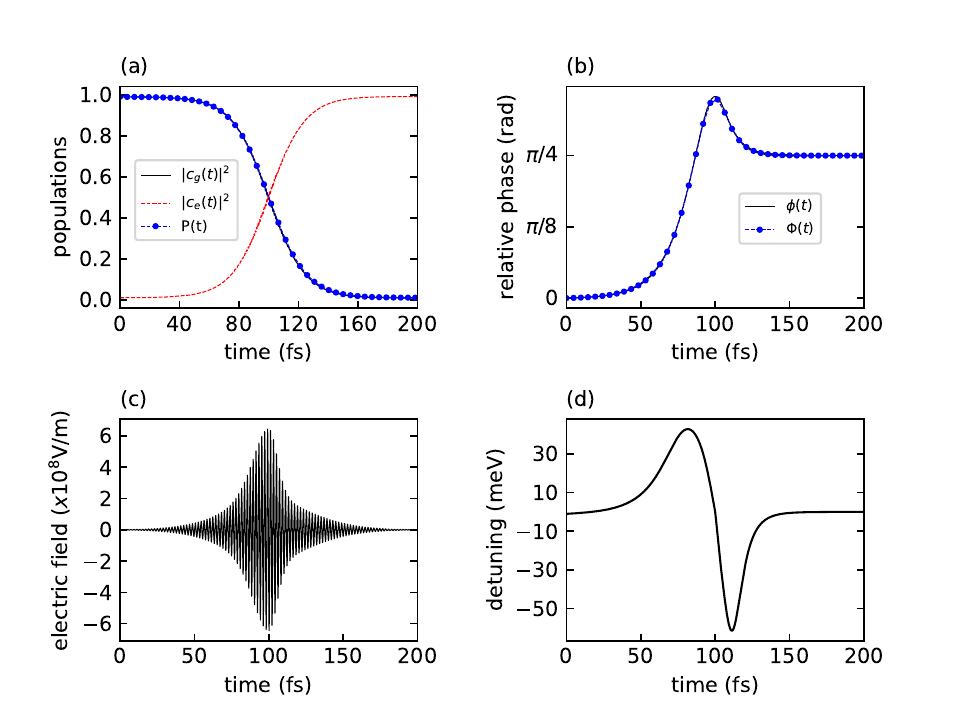}
\centering
 \caption{$P(t)$ is a hyperbolic tangent function joining asymptotically $P^i=0.99$ to $P^f=0.01$ and such that $P(\tilde{t})=1/2$ for $\tilde{t}=100$fs, see Eq.~(\ref{Ptanh}). The initial relative phase is $\Phi^i=0$ and the target phase is $\Phi^f=\pi/4$. A hyperbolic secant function is chosen for the relative phase such that $\Phi(t_0)=\Phi^i$, $\Phi(t_f)=\Phi^f$ and $\dot{\Phi}(\tilde{t})=0$, see Eq.~(\ref{sechPhi}). (a) Levels population dynamics along with $P(t)$; (b) Relative phase dynamics $\phi(t)$ (continuous line) compared to the chosen dynamics function $\Phi(t)$ (dotted line with points); (c) Applied electric field, Eq.~(\ref{Controlfield}), and (d) Detuning from the resonance frequency $\dot{\Phi}(t)+\dot{\Lambda}(t)$.}
\label{fig:TanhpopSechphase}
\end{figure}

In the previous applications, the pulses generated by the dynamical functions have the inconvenience of abrupt turning on and turning off. In order to design a smooth switch on and off of the pulses, we must chose both $\dot{P}(t)$ and $\dot{\Phi}(t)$ close to zero at $t_0$ and $t_f$ and with slow increasing/decreasing around these times. In order to illustrate this scenario, we chose for $\Phi(t)$ the following functional form which combines two hyperbolic secant functions,

\begin{align}\label{sechPhi}
\Phi(t)=
    \begin{cases}
        \chi_1\sech[\eta_1(t-\tilde{t})]+\sigma_1 & \text{, if } t<\tilde{t} \\
        \chi_2\sech[\eta_2(t-\tilde{t})]+\sigma_2 & \text{, if } t\ge \tilde{t}
    \end{cases},
\end{align}
where $\chi_1=(\Phi^{\rm max}-\Phi^i)/(\sech(\eta_1 \tilde{t})-1)$, $\chi_2=(\Phi^{\rm max}-\Phi^f)/(\sech[\eta_1 (t_f-\tilde{t})]-1)$, $\sigma_1=\Phi^{\rm max}-\chi_1$ and $\sigma_2=\Phi^{\rm max}-\chi_2$. The parameters $\eta_1$, $\Phi^{\rm max}$ as well as $\tilde{t}$ are freely chosen. $\Phi^{\rm max}$ sets the maximum value of $\Phi(t)$ at $t=\tilde{t}$, while $\eta_1$ relates to how fast this maximum is approached from the left. Furthermore, the relation $\eta_2=\eta_1\sqrt{\chi_1/\chi_2}$ guarantees the continuity of the second derivative of $\Phi(t)$ at $t=\tilde{t}$. Figure \ref{fig:TanhpopSechphase} presents the results with this choice of Eq.~$(\ref{sechPhi})$ for the dynamical function $\Phi(t)$ and Eq.~(\ref{Ptanh}) for $P(t)$. We have selected $P^i=0.99$ and $P^f=0.01$, while $\alpha=0.04{\rm fs}^{-1}$, $t_f=200$fs and $\tilde{t}=100$fs. For the relative phase, we choose $\Phi^i=0$ and $\Phi^f=\pi/4$. The parameters in Eq.~(\ref{sechPhi}) have been set to $\Phi^{\rm max}=1.4\Phi^f$ and $\eta_1=1.65\alpha$. Panels (a) and (b) shows that the population and relative phase follow the prescribed dynamics. Panel (c) shows that the control pulse has the desired bell-shaped envelope. Panel (d) shows that the detuning has a change in sign around the peak of the pulse.

\begin{figure}[h]
\includegraphics[width=15cm]{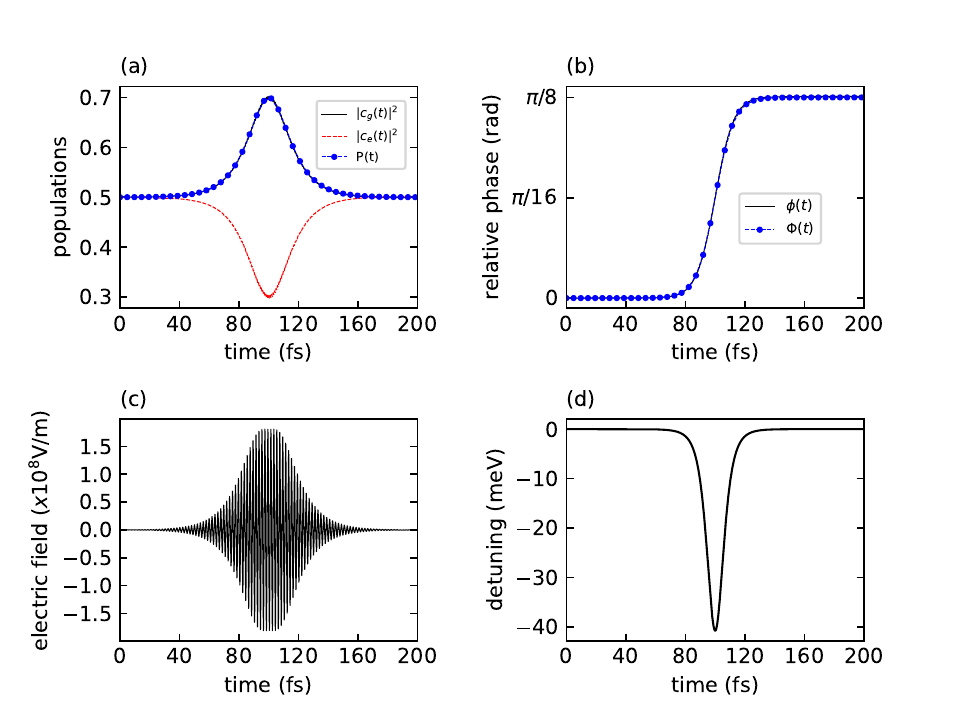}
\centering
 \caption{$P(t)$ is a hyperbolic secant function joining asymptotically $P^i=0.5$ to $P^f=0.5$ and such that $P^{\rm max}=0.7$ for $t_p=100$fs, see Eq.~(\ref{popsech}). The initial relative phase is $\Phi^i=0$ and the target phase is $\Phi^f=\pi/8$. A hyperbolic tangent function is chosen for the relative phase joining asymptotically $\Phi^i$ to $\Phi^f$ and such that $\Phi(t_p)=(\Phi^f-\Phi^i)/2$, see Eq.~(\ref{phasetanh}). (a) Levels population dynamics along with $P(t)$; (b) Relative phase dynamics $\phi(t)$ (continuous line) compared to the chosen dynamics function $\Phi(t)$ (dotted line with points); (c) Applied electric field, Eq.~(\ref{Controlfield}), and (d) Detuning from the resonance frequency $\dot{\Phi}(t)+\dot{\Lambda}(t)$.}
\label{fig:SechpopTanhchphase}
\end{figure}

Finally, we return to the problem of having equal initial and final populations, while changing the relative phase. As commented before, it is not possible to keep the populations constant during the hole time interval if they are equal or very close to $1/2$. However, it is possible to set $P^i=P^f=1/2$, without $P(t)$ being a constant. We address this situation choosing the hyperbolic secant function for $P(t)$,

\begin{equation}\label{popsech}
    P(t)=G\sech[\xi(t-t_p)]+F,
\end{equation}
where $G=P^{\rm max}-P^i$, $F=P^i$ and with $t_p=(t_f-t_0)/2$. The parameter $\xi$ sets how sharp is the function, while $P^{\rm max}=P(t_p)$ defines its maximum/minimum value, that can be arbitrarily chosen in the interval $0<P^{\rm max}<1$, except for $1/2$. Thus, the function $P(t)$ connects the same initial and final values. Note that since we will set $P^i=P^f=1/2$, the time derivative of $\Phi(t)$ does not need to vanish in the interval $t_0<t<t_f$. Therefore, we select a hyperbolic tangent function for $\Phi(t)$ joining asymptotically $\Phi^i$ to $\Phi^f$,

\begin{equation}\label{phasetanh}
    \Phi(t)=G\tanh[\chi(t-t^*)]+F,
\end{equation}
where $G=(\Phi^f-\Phi^i)/2$ and $F=(\Phi^f+\Phi^i)/2$. The parameter $\chi$ controls the transition time between the initial and final relative phase and $t^*$ is such that $P(t^*)=F$. We set $t^*=t_p$, which produces a pulse with an apparent symmetry around $t_p$. Figure \ref{fig:SechpopTanhchphase} illustrates the application of this scheme with the following set of parameters: $\Phi^i=0$, $\Phi^f=\pi/8$, $P^{\rm max}=0.7$, $t_f=200$fs, $t_p=t_f/2$ and $\xi=\chi=0.08{\rm fs}^{-1}$. As in the previous cases, panels (a) and (b) shows that the dynamics follows the prescribed dynamical functions. The envelope of the generated field in panel (c) has the expected bell-shape with a small plateau around $t=t_p$. The detuning presented in panel (d) indicates that the modulation of the frequency of the control field consists of adding a constant frequency to $\omega_0$, then there occurs an abrupt change around the time of the transition to a new constant frequency.

\section{Conclusion}
In this work, we have derived an analytic expression for the external field acting on a two-level quantum system and aimed at controlling simultaneously the population and the relative phase dynamics. The control field is engineered from two user-defined dynamical functions $P(t)$ and $\Phi(t)$ which join initial conditions to the target values. These functions must satisfy a series of conditions, but there is still great flexibility in their choices. It is worth emphasizing that the obtained expression can be easily build, needing, apart from the dynamical function themselves, only their derivatives. We have applied our approach to several initial and target values of population and relative phase, utilizing different dynamical functions. Our calculations show the general applicability of the approach, the only essential limitation being the validity of the RWA. We expect the present result to be used beyond the simple two-level system, e.g., in the strong-field control and for the control of many-level molecular systems \cite{PhysRevA.106.043113,PhysRevA.104.063102}.

\section*{Acknowledgments}\label{sec7}
EFL acknowledges support from the Brazilian agency S\~ao Paulo Research Foundation, FAPESP (grants 2023/04930-4 and 2014/23648-9). FSF and EFL acknowledge support from National Council for Scientific and Technological Development - CNPq-Brazil through the PIBIC-Scholarship 2023-2024.

\pagebreak

\pagebreak

\pagebreak

\pagebreak

\pagebreak

\pagebreak

\pagebreak


%

\end{document}